# Agent-based (BDI) modeling for automation of penetration testing

Ge Chu
Department of Computer Science
University of Liverpool
Liverpool, UK
gechu@liverpool.ac.uk

Alexei Lisitsa
Department of Computer Science
University of Liverpool
Liverpool, UK
lisitsa@liverpool.ac.uk

*Abstract*—Penetration testing (or pentesting) is one of the widely used and important methodologies to assess the security of computer systems and networks. Traditional pentesting relies on the domain expert knowledge and requires considerable human effort all of which incurs a high cost. The automation can significantly improve the efficiency, availability and lower the cost of penetration testing. Existing approaches to the automation include those which map vulnerability scanner results to the corresponding exploit tools, and those addressing the pentesting as a planning problem expressed in terms of attack graphs. Due to mainly non-interactive processing, such solutions can deal effectively only with static and simple targets. In this paper, we propose an automated penetration testing approach based on the belief-desire-intention (BDI) agent model, which is central in the research on agent-based processing in that it deals interactively with dynamic, uncertain and complex environments. Penetration testing actions are defined as a series of BDI plans and the BDI reasoning cycle is used to represent the penetration testing process. The model is extensible and new plans can be added, once they have been elicited from the human experts. We report on the results of testing of proof of concept BDI-based penetration testing tool in the simulated environment.

## I. Introduction

In recent years, malicious network attacks have become an increasingly serious threat to individuals, businesses and even national information security. Penetration testing [1] is a methodology which simulates real attacks with the aim to assess the security of computer systems and networks. The main distinction between an attacker and penetration testing depends on the legality. In other words, penetration testing aims to improve the security of the system rather than destroy or access information illegally and it does not affect the availability of target systems. The process of penetration testing is normally done manually, and the test cycle is relatively long. Moreover, the test results are highly dependent on the level of skill and experience of a tester or penetration team. To improve the efficiency, automated penetration testing methods and tools are needed. The automation can significantly reduce the time, cost and human involvement in the process of information gathering, analysis and exploitation.

Existing approaches to the automation include those mapping vulnerability scanners results to the corresponding exploitation tools, and those addressing the pentesting as a planning problem expressed in terms of attack graphs. Due to mainly non-interactive processing, such solutions can deal effectively only with static and simple targets. However, the target environment of penetration testing is normally dynamic, uncertain and complex. The human penetration tester needs to interact with the environment or target and choose the best action to compromise the target system based on the feedback and their interpretation. In order to deal with these issues, we propose to use an agent-based architecture for the automation of pentesting. An agent [2] can interact with the environment by perception, decision making and action. Moreover, the behavior of an agent can be flexible and can be generally characterized as autonomous, reactive, proactive and social. Currently, agent-based technologies are considered as promising for the applications in various areas. There are three main kinds of agent architectures considered in the literature, these are Reactive, Cognitive and Hybrid [3]. The BDI agents (Belief-Desire-Intention) is one of the classical and most representative models of Cognitive architecture which is proposed by Bradman [4]. The BDI model enables agents to have cognitive abilities to deal with dynamic, uncertain and complex environments by allowing for mental states, characteristics/attitudes such as belief, desire and intention.

In this paper, we propose an agent-based BDI model with the aim to improve the efficiency and probability of success for automated penetration testing. Penetration testing actions are defined as a series of BDI plans and the BDI reasoning cycle is used to represent the penetration testing process. To validate this model, we implement a prototype system and have simulated real world penetration testing scenarios using agent-based programming language Jason [5]. The rest of this paper is organized as follows: Section II introduces the related work on automated penetration testing. In section III, we propose the agent-based BDI model for penetration testing. In section IV, we present an implementation of a prototype of our BDI model. In section V, we present the experiment and validate our model. Finally, we close this paper with a conclusion and further work in section VI.

## II. RELATED WORK

Xue Qiu et al. [6] proposed an automated method of penetration testing named AEPT (automata model of penetration testing) based on a four-stage model of penetration testing. They defined testing time, target, scheme, plan, the collection of scanning information in addition to the analysis of exploiting vulnerabilities. This model was then subsequently used to generate the penetration testing scheme automatically. Finally, they proposed the automatic executing method of penetration testing scheme by calling the exploitation module. However, AEPT only tries to exploit all vulnerabilities of the target after receiving the scanning report as an input and fails to take into account the dynamic and uncertain nature of a situation in a real-world penetration testing scenario. In addition, a real penetration tester often attempts to compromise a target via multi-step attacks using a series of exploitation tools, in particular to recover from failed attempts. AEPT is unable to deal with such attack chain situations.

The majority of approaches to date address automated penetration testing as a planning problem for an attack graph. Cynthia and Swiler [7] presented a graph-based flexible approach to perform system vulnerability analyses. This analysis system broken database of common attack into atomic steps, specific network configuration, topology information and attacker profile. Nodes and arcs in the attack graph represent the stage of an attack. The probabilities of success will be assigned to the arcs and various graph algorithms was applied to identify the attack paths with the highest probability of success. Kyle et al. [8] created a NetSPA attack graph system which allows network defenders to evaluate threats and choose corresponding countermeasures. NetSPA is able to analyze numerous targets within a few minutes by using firewall rules and vulnerability scans. Moreover, asset values are assigned to each target in order to measure the purpose or mission. Xue Qiu et al. [9] proposed an automatic generation algorithm of penetration graph that makes use of CVSS (Common Vulnerability Scoring System) to increase the reliability of attack paths, which ultimately optimizes the network topology. This algorithm probes and represents the network topology by matrix and searches the path to the target, which can generate the attack graph from the vulnerability scanner result. However, the limitation of the aforementioned graph-based methods is that they can only output the action sequence to deal with stationary environment. This is turn can only provide the steps or guidelines for penetration testing therefore they are still unable to perform interactively within real world penetration testing scenarios.

There have been numerous applications of agent-based models in the real world such as in agriculture, air traffic control, economics, emergency evacuation, healthcare and social behavior [10]. The penetration testing scenarios mentioned above are similar because of the dynamics, the uncertainty, interactivity and complexity of the environment. There is not, as of yet, an approach which would able to deal with the above characteristics in automated penetration testing scenarios. Therefore, we propose an agent-based BDI model to achieve automated penetration testing with high efficiency and a higher probability of success.

## III. AGENT-BASED BDI MODEL FOR PENETRATION TESTING

In penetration testing, humans need to create the goals and plans to obtain a successful result. Agent-based BDI is a natural candidate to model this problem because it can interact with the target and performs various types of attacks. In this section, we discuss how to model penetration testing problems using an agent-based BDI model.

In the process of penetration testing, the BDI agent interacts with the target by perceiving information and in response it outputs actions to change it. In our model, we only consider the single agent situation, but the number of targets is unlimited. The agent world consists of the network environment such as the Internet or the local area network and we assume that the agent can interact with targets via different kinds of connections either wired or wireless. In the action space, we pre-define different types of actions to be performed throughout the whole penetration testing process from the information gathering stage to the report stage. Whereas some scanners or penetration testing tools provide a degree of automation, our model can execute external tools directly as part of the action space in order to make this model more extensible. Moreover, to compare with other approaches or tools, our model can perform various types of attacks such as buffer overflow attack, SQL injection attack, password attack, sniffer attack and social engineering attack.

The BDI model defines the process of an agent choosing actions according to target information in penetration testing. The basic logic components of a BDI agent are belief, desire and intention. In our model we follow the conventions adopted in the Jason Interpreter, which in turn are based on PRS (Procedural Reasoning System) [5].

BDI agent is defined as a tuple <Ag, B, D, I, P, A, S>, where Ag is an agent name, B is a belief set, D is a desire set, I is an intention set, P is a plan set, A is an action set and S is a Perception set. Now we explain all components of this definition.

*Belief set* B represents the set of information about the target and it will be updated after executing actions. In the context of pentesting, this kind of information typically comprises OS type, open port, DNS, service name or version, vulnerability, configuration, network topology and privilege, etc. New beliefs will be generated based on current belief and perceived information.

$$B = f1(B \times S) \qquad (1)$$

*Desire set* D represents all the options or possible candidate plans of penetration testing for the agent that might like to accomplish. In real time penetration testing, multiple kinds of attack methods can be carried in response to specific target information. For example, SQL injection attack, password attack or buffer overflow attack can be carried out when the target port 80 is opened and human penetration testers would need to choose one type of attack according to their experience/preferences. The desire is determined based on beliefs and intentions.

$$D = f2(B \times I) \quad (2)$$

*Intention set* I represents the agent goals or which plan the agent decides to carry out. In penetration testing, the agent needs to choose one plan to carry out from the possible candidate plans. Namely, the plan becomes intention after being selected.

$$I = f3(B \times D \times I) \quad (3)$$

*Plan set* P consists of available plans, each giving the information about how to achieve the goals. A plan comprises three parts: trigger event, context and body. The trigger event is an event that the plan can handle such as beliefs or goals. The structure of the plan is shown in Figure 3: The context defines the prerequisites under which the plan can be used. The body defines a series of actions to be carried out if the plan is chosen. In our model, we pre-define various types of information gathering actions and attack methods.

Trigger Event: context <- body.

The BDI agent reasoning cycle for penetration testing is described below:

1. Initial beliefs and intentions will be set up by the penetration tester and normally represents information regarding the target such as the domain or IP address and the privilege which the penetration testing must achieve, respectively.

2. The BDI agent perceives the target information by performing various information gathering actions. For example, Nmap can collect OS type and ports opened at the target.

3. After perceiving the feedback, current beliefs will be updated. At this time, the BDI agent should hold the current information about the target.

4. According to the new current belief, all relevant action plans will be found. For example, if port 80 of the target is opened, then password attack, buffer overflow attack, SQL injection attack are all become candidate options for the human penetration tester.

5. The BDI agent chooses one plan from the candidate action plans to become the intention and waits to be executed according to the context of the plan and the human knowledge database which chooses the plan based on human penetration testing experience in the real world. We pre-define the priority of the chosen actions in the human knowledge database.

6. The BDI agent executes the chosen plan. If the plan fails, then the agent chooses another plan.

7. The BDI agent checks whether the initial goal is achieved or not and decides either (1) to output the report which records the process of the whole penetration testing or (2) to return back to the new reasoning cycle.

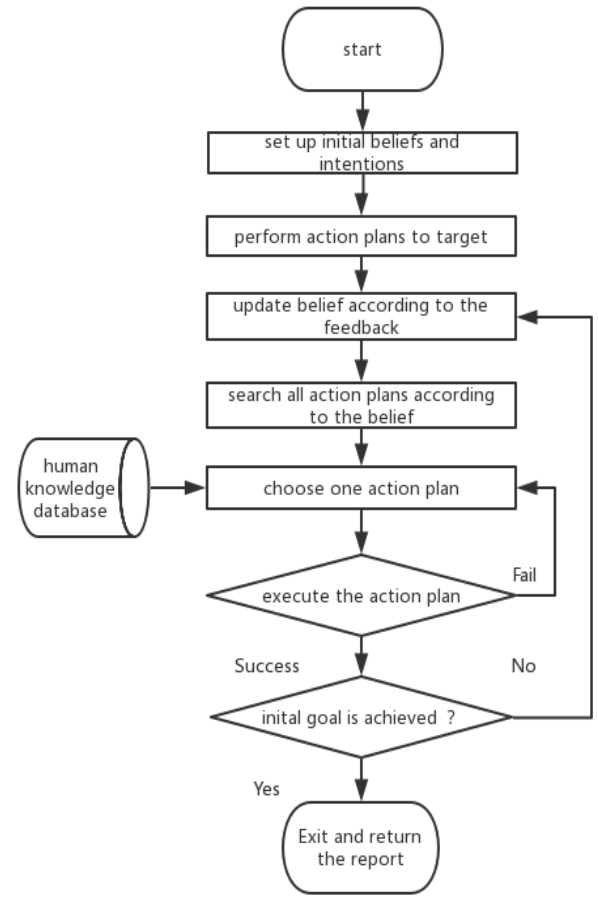

Figure 3 The BDI agent reasoning cycle for penetration testing

IV. IMPLEMENTATION OF BDI FOR PENETRATION TESTING

Our model is implemented in AgentSpeak Jason which is a multi-agent system programming language based on belief-desire-intention paradigm (BDI). The reason we implemented our model in Jason is that Jason is one of the best known and well-established agent-based development languages for cognitive agents. Jason takes it origins from the Procedural Reasoning System (PRS) developed at SRI in late 1980s [5]. What is more, Jason is implemented in Java (running in multi-platform) and provides interfaces to call Java code, which enables our model to use external tools. We pre-define various actions to cover whole the penetration testing stage from information gathering to report.

### A. Information gathering:

The first stage of penetration testing is information gathering. In our model, the scanner or the various information gathering tools are used to probe the target and update the belief. We use Nmap to collect OS type, IP address, port open, services information from the target. Openvas and Nessus are mainly used to probe the OS vulnerability such as bufferflow, configuration, and information leakage. The Harvester is used to collect email addresses to perform social engineering attacks.

### B. Buffer overflow attack

After the information gathering stage, the buffer overflow attack will be exploited as the preference if there are remote buffer overflow vulnerabilities found and updated in the belief set. We make use of metasploit which is the most critically acclaimed and widely used penetration testing framework to perform buffer overflow attack in our model due to it having the ability to collect thousands of exploit codes to attack various OS.

### C. Sql injection attack

If the target is running a web server, our model will try to perform an Sql injection attack. After the information gathering stage, the Sql injection vulnerabilities will be listed in the scanner. In our model, we make use of W3af and SQLmap to probe and perform the SQL injection attack. After the SQL injection attack, the web privilege is obtained and the model will perform further actions to improve privilege.

### D. Password attack

The Password attack will be performed if there are services allowing users to log in remotely such as ssh, ftp, Telnet and SQL database, etc. The dictionary will be generated according to information about the target and will perform an attack by Hydra. We will obtain access privilege if this type of attack is successful. Nevertheless, the successful rate is nominally low and a time-consuming activity.

### E. Sniffer attack

Aforementioned attacks are not successful if there is any vulnerability in the well protected target. In these situations, human experienced penetration tester would normally attempt to break into another system which is under the same subnetwork with the original target and perform the sniffer attack or Man in the Middle attack to obtain access privilege on it. In our model, the Arpspoof and Ettercap will perform these types of attacks respectively.

### F. Social engineering attack

Setoolkit used to accomplish social engineering attacks such as spear phishing attack, web forge attack and powershell attack. These kinds of attacks are normal to humans such as administrator or target system staff which have weak security awareness by sending deliberately structured emails to the target administrator or staff to obtain access privilege directly.

### G. Report generation

Our model will record the attack action name and path to show the process of the whole penetration testing and output it as report if a plan is executed successfully. We use the internal action of Jason to achieve this function.

```
1   ip_address(192.168.0.10). // initial belief
2   !privilege(root). //initial goal
3
4   // plans
5   @basic information gathering
6   +! get(port) : true
7   <- !Nmap(ip_address).
8   .print(Namp ip_address).
9
10  +! get(os) : true
11  <-  !Nmap(ip_address);
12  .print(Namp os).
13
14  +!get(service) : true
15  <- !Nmap(ip_address);
16  .print(Namp service).
17
18  @vulnerability information gathering
19  + get(vulnerability) :   true
20  <- !openvas(ip_address);
21  .print(openvas ip_address).
22
23  @ Buffer overflow attack
24  +get(vulnerability)
25  : get(vulnerability) != null
26  <- !Metasploit(cve no.);
27  .print(metasploit ip_address).
28
29  @Sql injection attack
30  +get(port == 80)
31  : get(service == apache or nginx)
32  <- !SQLmap(ip_address);
33  .print(SQLmap ip_address).
```

Figure 4 A part of Jason code for penetration testing

Figure 4 shows a part of the Jason code based on the BDI model for penetration testing. Firstly, we set up the initial belief and initial goal of our BDI agent as ip address and root privilege. Then we pre-define the basic information gathering plan to probe the target Opened port, OS type and application services information by Nmap. After the basic information gathering stage, we perform the vulnerability information gathering by openvas and perform the buffer overflow attack using metasploit or an SQL Injection attack by SQLmap if the per-conditional is satisfied. Each of these actions will be recorded by Jason's internal action named print function.

## V. EXPERIMENT

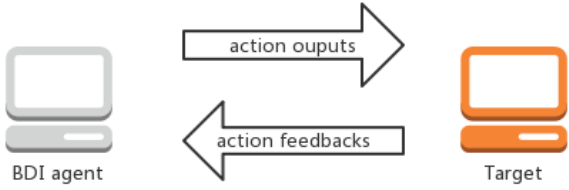

Figure 5 The interaction between BDI agent and Target

Our model runs on a PC with an Intel I7 CPU at 2.0 GHz and 4GB of RAM. As we can see in figure 5, The simulation experiment consists of two agents to represent the BDI agent and the target. In order to simplify the process of penetration testing in the virtual environment, we use the internal communication actions in Jason to simulate the interaction between the BDI model and the target agent.

### A. Target agent

| OS | Port | Services | Vulnerability | Password |
|---|---|---|---|---|
| Linux | 80, 22, 3306 | Nginx, SSH, MySQL | CVE-remote CVE-local | SSH:456 |

Table II    Target information

We set up basic information regarding the target including the system type, opened port, service, vulnerability and the SSH password in the initial belief set to simulate a target server as shown in Table II. To make the scenario uncertain, we use randomization and set 0.8 as the threshold to determine if the SSH password attack is successful by generating a random number and comparing it with the threshold. In terms of the remote or local buffer overflow attack successful rate, we set thresholds as 0.5 and 0.3, respectively (this is based on personal penetration testing experience of the first author)

### B. BDI agent

In BDI agent, we set up the value of privilege as none initially and the initial goal is root privilege. We pre-define information gathering plans to probe OS type, opened port, service and vulnerability information from the target agent. To simplify the process of penetration testing by the BDI agent, we pre-define the password attack and buffer overflow attack to target.

### C. Reasoning process between BDI agent and target agent

We carry out two simulations to show how our BDI agent can perform in different circumstances in below:

*1) Simulation 1*

Figure 6 BDI agent result in simulation 1

We can see from the output of the processes of the BDI agent in Figure 6, the BDI agent probed all information about the target in the belief set but failed to perform the password attack because the rate of the password attack has not reached the specified 0.8 threshold. Hence, the BDI agent cannot perform local the buffer overflow attack as well since we define the prerequisite of it as successful password attack. However, the remote buffer overflow attack was successful and the current privilege has changed to root. We can check the validity of the process in Figure 7.

Future 7 BDI agent Belief set in simulation1

*2) Simulation 2*

In this simulation, the BDI agent probed all the information of the target and successfully broke the SSH password because the rate of the SSH password attack was set to 0.9 which is greater than the 0.8 threshold. Moreover, the BDI agent performed successfully in both the local and the remote buffer overflow attacks. We can see the privilege change from none to the user then reached to root in Figure 8.

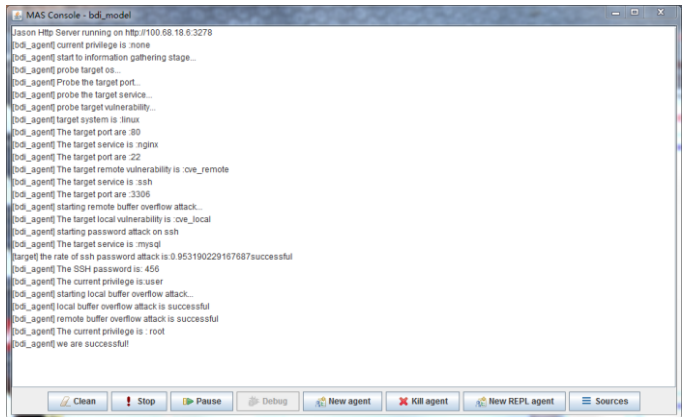

Figure 6 BDI agent result in simulation 2

VI. CONCLUSION

This paper presents an agent-based Belief-Desire-Intention(BDI) modelling for the automation of penetration testing, which enables interaction between dynamic and uncertain targets. Penetration testing actions are defined as a series of BDI plans and the BDI reasoning cycle is used to represent the penetration testing process. Two simulations show the BDI agent behavior and reasoning process to validate the modelling. Our current and future research aims to extend the model with more types of actions to deal with complex real-world pentesting scenarios and to experiment with real (non-simulated) environments.